\documentclass[notitlepage,onecolumn]{revtex4-1} 

\usepackage{hyperref}   
\usepackage{amsmath}    
\usepackage{graphicx}                    
\usepackage{epsfig} 
\usepackage{epstopdf}

\begin{document}

\title{
Master equation for operational amplifiers:
stability of negative differential converters,
crossover frequency and pass-bandwidth}

\author{T.~M.~Mishonov, V.~I.~Danchev, E.~G.~Petkov}
\email[E-mail: ]{mishonov@bgphysics.eu}
\affiliation{Faculty of Physics, St.~Clement of Ohrid University at Sofia,\\
5 James Bourchier Blvd., BG-1164 Sofia, Bulgaria}

\author{V.~N.~Gourev}
\affiliation{Department of Atomic Physics, Faculty of Physics,\\St.~Clement of Ohrid University at Sofia,\\5 James Bourchier Blvd., BG-1164 Sofia, Bulgaria}

\author{I.~M.~Dimitrova} 
\affiliation{Faculty of Chemical Technologies, University of Chemical Technology and Metallurgy, \\
8 Kliment Ohridski Blvd., BG-1756 Sofia}

\author{N.~S.~Serafimov}
\affiliation{Faculty of Telecommunications, Technical University Sofia, \\8 Kliment Ohridski Blvd., BG-1000 Sofia, Bulgaria}

\author{A.~A.~Stefanov}
\affiliation{Faculty of Mathematics, St.~Clement of Ohrid University at Sofia,\\
5 James Bourchier Blvd., BG-1164 Sofia, Bulgaria}

\author{A.~M.~Varonov}
\email[E-mail: ]{avaronov@phys.uni-sofia.bg}
\affiliation{Department of Theoretical Physics, Faculty of Physics,\\
St.~Clement of Ohrid University at Sofia,\\
5 James Bourchier Blvd., BG-1164 Sofia, Bulgaria\\}

\date{07 April 2019}

\begin{abstract}
The time dependent master equation from the seminal article by
Ragazzini, Randall and Russell
[J.~R.~Ragazzini, R.~H.~Randall and F.~A.~Russell,
``Analysis of Problems in Dynamics by Electronic Circuits'',
Proc.~of the I.R.E., \textbf{35}, pp.~444--452, (1947)] is recovered as necessary tool
for the analysis of contemporary circuits with operational amplifiers.
This equation gives the relation between time dependent the output voltage $U_0(t)$ 
and the difference between the input voltages ($U_{+}(t)$ and $U_{-}(t)$). 
The crossover frequency $f_0$ is represented the time constant $\tau_{_0}$ of this equation.
The work of the master equation is illustrated by two typical examples:
a) the stability criterion of the devices with negative impedance converters, 
which we consider as a new result
b) the frequency dependence of the amplifiers with operational amplifiers given in the 
technical specifications without citations of time dependent equation.
A simple circuit for determination of $f_0$ is suggested and the method is illustrated 
by determination of crossover frequency for 
the low-noise and high speed ADA4898 operational amplifier.
It is concluded that
for an exact calculation of the pass bandwidth of amplifiers with active filters
the 70 years old master equation is a useful technique implicitly included in the contemporary software.
The frequency dependent formulae for the amplification coefficient of inverting and non-inverting amplifiers are given for the case of non-zero conductivity between the inputs of the operational amplifiers. 
\end{abstract}

\maketitle

\section{Introduction}

Inspired by Heaviside operational calculus 70 years ago, Ragazzini, Randall and Russell~\cite{Ragazzini:47} introduced the idea and coined the generic term operational amplifier~\cite{Jung-h:02}
 (enabling summation, integration and differentiation), 
 and introduced the master equation 
\begin{equation}
U_+ - U_- = G^{-1} U_0,
\quad
\hat{G}^{-1}=\frac{1}{G_0}+\tau_{_0}\hat{s},
\label{master}
\end{equation}
which describes the relation between input voltages $U_+(t)$ and $U_-(t)$ and output voltage $U_0(t)$
Here $\hat s$ is the time differentiating operator 
\begin{equation}
\hat s=\frac{\mathrm{d}}{\mathrm{d}t},
\end{equation}
which for exponential time dependence of the voltages $U\propto \mathrm{e}^{st}$ is reduced to its eigenvalue $\hat s \mathrm{e}^{st}=s\mathrm{e}^{st},$ and for the reciprocal
open loop-gain we have 
\begin{equation}
G^{-1}(\omega)=G_0^{-1}+s\tau_{_0}, \quad s=\mathrm{j}\omega,
\label{eigen}
\end{equation}
where $\mathrm{j}^2=-1$ and $\omega$ is the frequency.
As static open-loop gain $G_0\gg1$ for all high frequency applications, the $G_0^{-1}$ term is negligible.
The time constant of the operational amplifier is represented by the crossover frequency $f_0$
\begin{equation}
\tau_{_0}=\frac1{2\pi f_0},
\end{equation}
called also -3dB frequency of the open-loop gain.
Implicitly the equation from the work of Ragazzini, Randall and Russell~\cite{Ragazzini:47} exists in Fourier representation in the modeling of operational amplifiers since 
1970~\cite{Budak:81,SPICE,Sun,Ghausi,Mohan,Rashid} 
but never as a differential equation describing time dependent voltages
\begin{equation}
U_+(t) - U_-(t) = \left(\frac{1}{G_0}+ \tau \frac{\mathrm{d}}{\mathrm{d}t}\right)U_0(t)
\label{TimeDependent}
\end{equation}
which is necessary for the transient analysis given by the commercial software.
The Fourier representation of this time dependent equation Eq.~(\ref{eigen}) can be fund however 
in many sources.
Here we wish to emphasize that fundamental equations of physics and engineering are given in time representation; let us recall Maxwell. Schoedinger and Newton equations.

Nevertheless, the master equation of the operational amplifier is in some sense new an it is almost impossible to find a description how the procedure for determination of $f_0$ can be derived and how it is possible to calculate the frequency dependence of the amplification of non-inverting amplifier, for example, given in the the many  technical applications, see for example the specification for operational amplifier ADA4817~\cite{ADA4817}.
The name of the equation itself is not even finalized cf. \cite{Motivation} and in many applications the frequency dependence is neglected leaving no trace of any time or frequency relation.
Another important example is the stability problem of circuits with operational amplifiers.

Instead of review, here we will present only two typical examples illustrating the applicability of the master equation Eq.~(\ref{master}) to the negative impedance converter (NIC) and 2) the determination of crossover frequency $f_0$ by frequency dependence of the amplification of a non-inverting amplifier analyzed in the 
next two sections.

\section{Negative impedance converter with operational amplifier}
 
Negative impedance convertors (NIC) are an important idea in electronics~\cite{Merill:51,Linvill:53,Pippard:78}
and they can be easily constructed by operational amplifiers.
\begin{figure*}[h]
\centering
\includegraphics[scale=0.4]{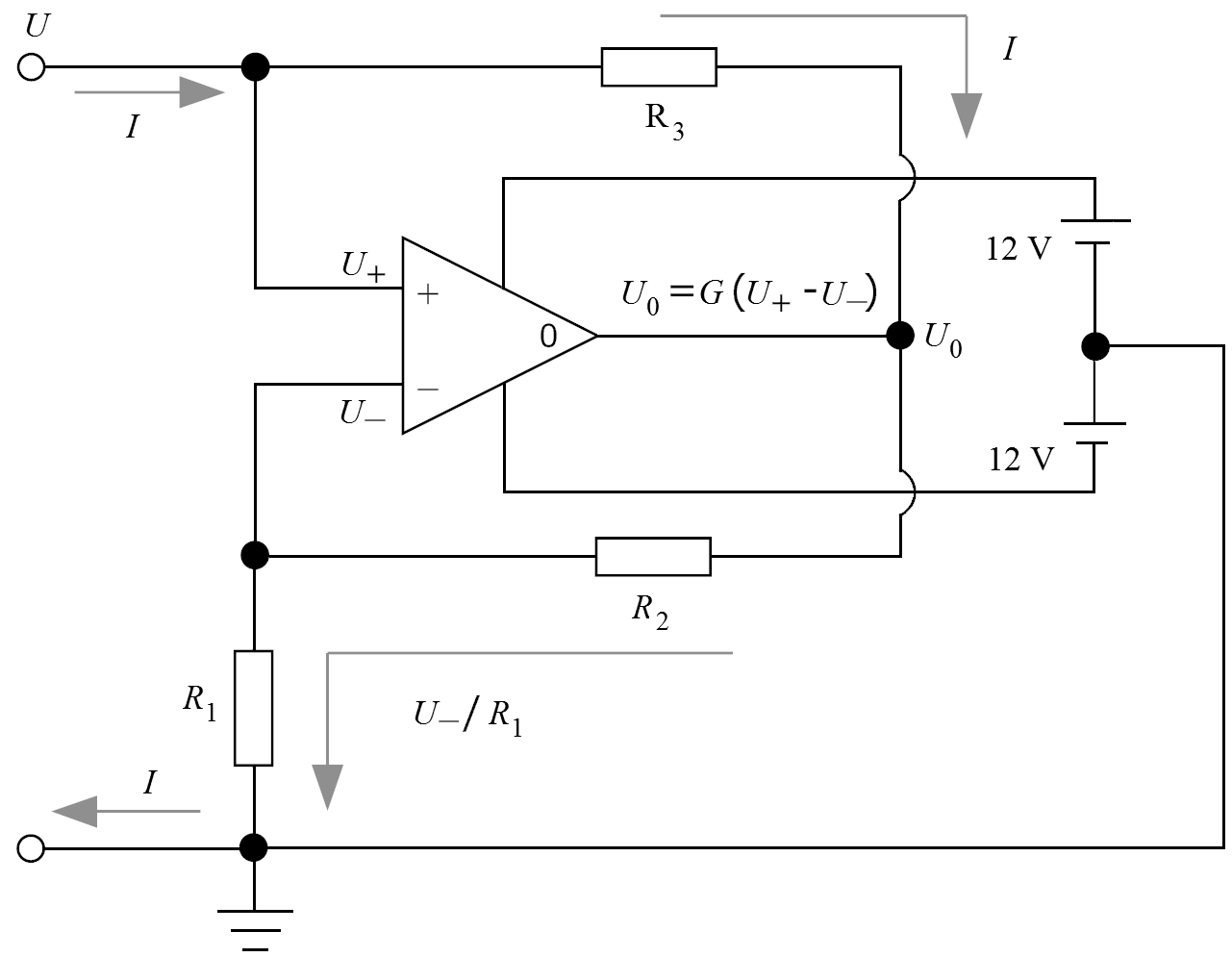}
\caption{Negative impedance converter built by operational amplifier according NASA patent~\cite{Deboo:68}.
For static voltages and negligible reciprocal open-loop gain $G_0^{-1}\ll1$ the circuit gives
for the ratio of input voltage and current $U/I=-R_1R_3/R_2.$
}
\label{ndc}
\end{figure*}

A circuit for NIC is depicted in Fig.~\ref{ndc}; the figure is taken from Ref.~\cite{EPO3}.
Solving the corresponding Ohm equations 
\begin{equation}
U-U_-=G^{-1}U_0,\quad I=\frac{U-U_0}{R_3}=\frac{U_0}{R_1+R_2}=\frac{U_-}{R_1}, 
\end{equation}
together with Eq.~(\ref{eigen})
after some high-school algebra we obtain the impedance of the circuit
\begin{equation}
Z\equiv\frac{U}{I}= \dfrac{R_3}{1-\dfrac{1}{\frac{R_1}{R_1+R_2}+G_0^{-1}+\tau_{_0} s}}.
\label{NegativeImpedanceConverter}
\end{equation}
Let us recall the derivation: 
1)~from the last equation we express $U_-=U_0 R_1/(R_1+R_2)$;
this is actually a voltage divider;
2)~then we substitute so expressed $U_-$ in the first equation and obtain
$U_0=U/\left[R_1/\left(R_1+R_2\right)+G^{-1}\right]$;
3)~finally we substitute $U_0$ in the first formula for the current $I$ and arrive at 
Eq.~(\ref{NegativeImpedanceConverter}).
In the static case $s=0$ and negligible $G_0^{-1}\ll1$ this equation gives
\begin{equation}
Z(\omega=0)\approx -\overline R,\quad \overline R\equiv\frac{R_1R_3}{R_2}.
\end{equation}
This result can be checked by current-voltage characteristics of the NIC.

\begin{figure}[h]
\centering
\includegraphics[scale=0.27]{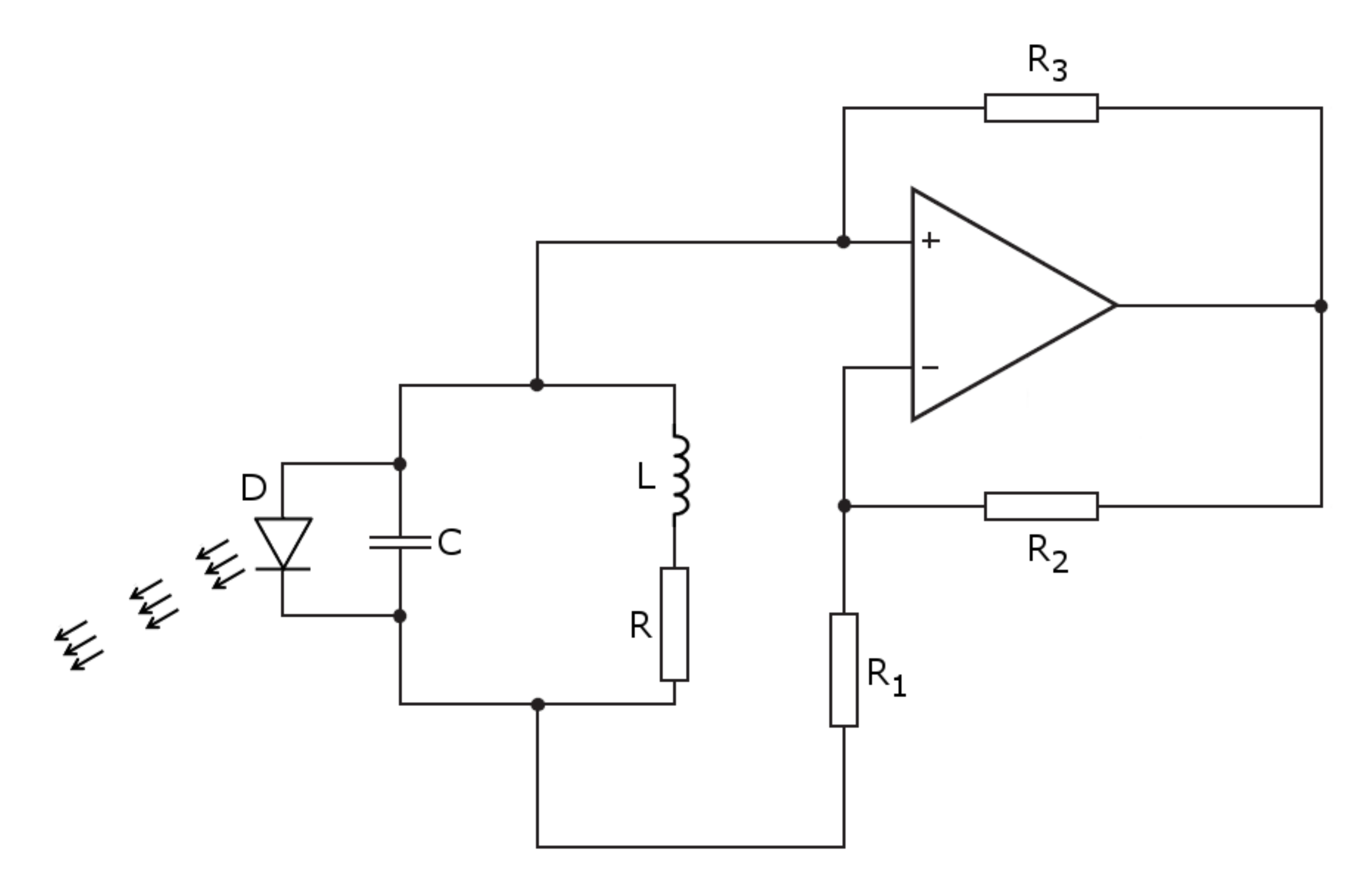}
\caption{Can a NIC excite electric oscillations in a parallel resonance circuit? 
Light emitting diode (LED) D radiates a pulsating light and limits the amplitude of the electric oscillations.}
\label{par}
\end{figure}

If an external load resistor $R_\mathrm{e}$ is connected in parallel to the NIC the total conductivity is zero
\begin{equation}
\frac1{Z}+\frac1{R_\mathrm{e}}=0
\end{equation}
and this eigenvalue problem has an explicit solution
\begin{equation}
\tau_{_0} s+ G_0^{-1}= \frac1{1+\frac{R_3}{R_\mathrm{e}}}
                              -\frac1{1+\frac{R_2}{R_1}},
\label{stab}
\end{equation}
for the real in this case exponent $s.$
For negligible $G_0^{-1}\ll 1$ $s$ is positive if $R_3/R_\mathrm{e}<R_2/R_1$ or
\begin{equation}
R_\mathrm{e}> \overline{R}\equiv R_1R_3/R_2.
\label{s0}
\end{equation}
Positive and real value of $s$ means exponential increase $\mathrm{e}^{st}$ of the voltage up to appearance of nonlinear effects, for example lightning of a light emitting diode (LED).
A voltmeter will show significant voltage and such an impedance cannot be measured by a digital Ohmeter.
Such circuit, however, can be used for generation of voltage oscillations in parallel resonance circuit with big resistance in the resonance shown in Fig.~\ref{par}. 
For high quality factor $Q=\sqrt{L/C}/R\gg1$ resonance contours the effective resistance 
$R_\mathrm{eff}=Q^2R>\overline{R}$ and in the circuit down in Fig.~\ref{par} electric oscillations are excited.
When an instability parameter $(R_\mathrm{eff}-\overline{R})/\overline{R}\ll1$ is relatively small,
the amplitude of the voltage oscillations is slightly above the forward voltage $U_c$
(the critical voltage to light the LED) of the LED and one can evaluate the consumed by the diode power as $P=U_c^2/2\overline{R}.$

\begin{figure}[h]
\centering
\includegraphics[scale=0.3]{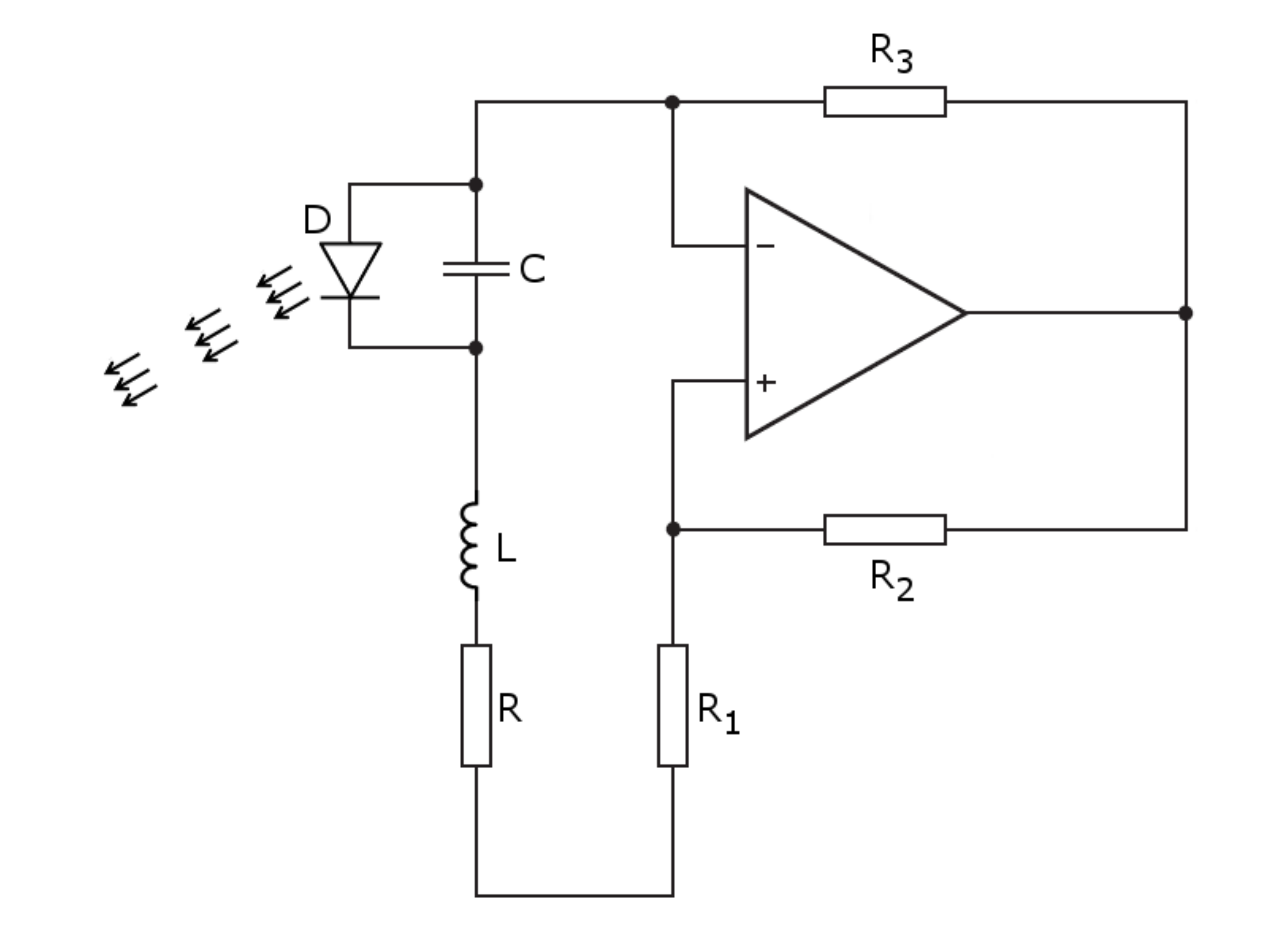}
\caption{Excitations of electric oscillations by NIC in a sequential resonance circuit. 
Why is the polarity of the operational amplifier opposite 
with the polarity of the parallel resonance circuit? 
Can polarities be calculated using the theoretical description of the work of operational amplifiers?
Search through the Internet.}
\label{seq}
\end{figure}

\begin{figure}[h]
\centering
\includegraphics[scale=0.3]{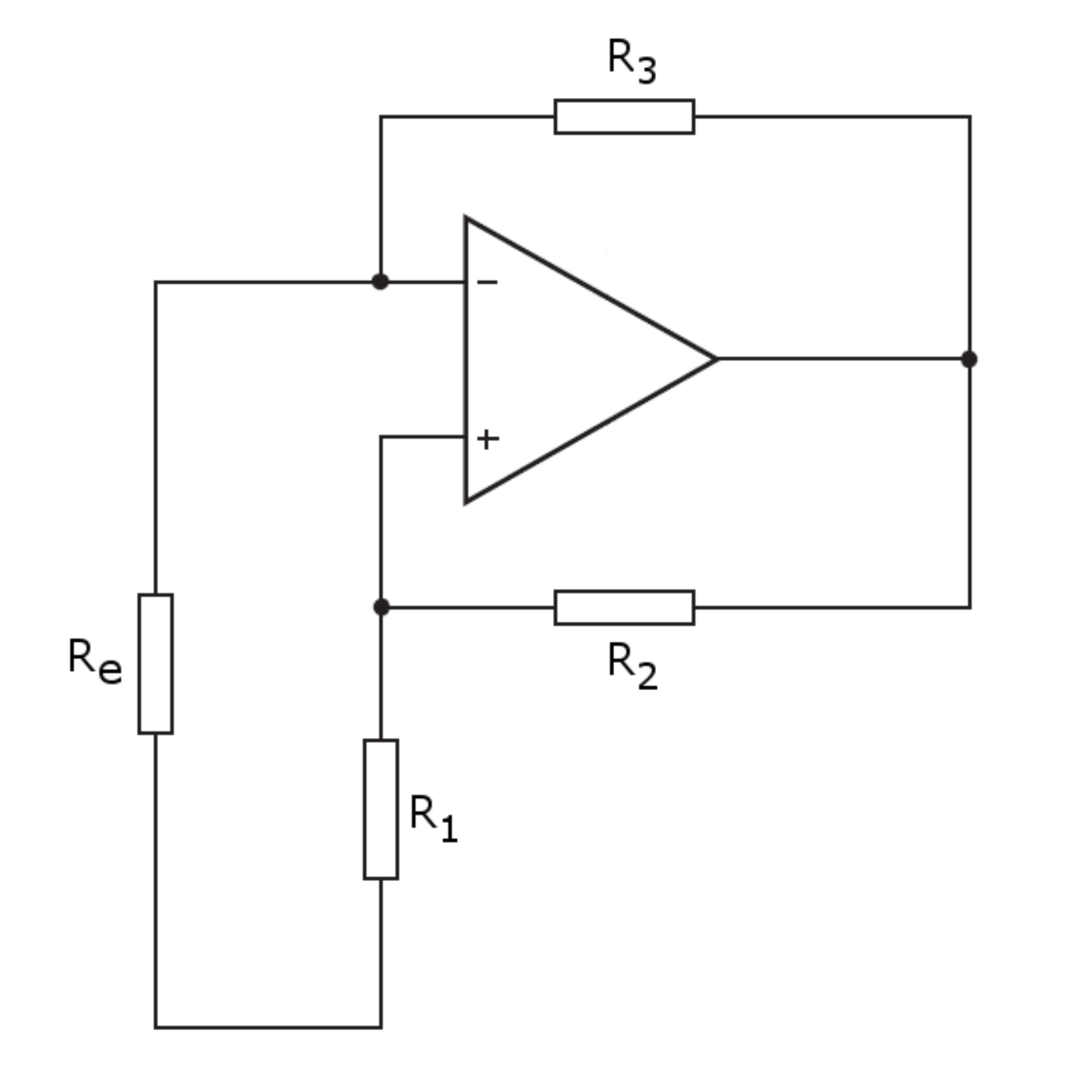}
\caption{Whether a digital Ohmmeter with big internal resistance $R_\mathrm{e}$ can show negative Ohms
$-\overline{R}$, where the parameter $\overline{R}\equiv R_1R_3/R_2.$}
\label{ohm}
\end{figure}

However, if we change the polarity of the operational amplifier as it is illustrated in 
Fig.~\ref{seq}, it leads formally to change of the sign in the stability equation
Eq.~(\ref{stab}). 
For such circuits NIC creates electric oscillations in sequential resonance circuits with small 
resistance of the coil $R\ll\sqrt{L/C}.$
For high resistance loads, however, the inverted circuit from Fig.~\ref{seq} is stable.
If we connect a voltmeter it will show zero voltage and some digital Ohmmeters can show negative resistance $-\overline{R}$ as it is demonstrated in Fig.~\ref{ohm}

Analyzing the work of NIC, we suppose that the exponent $s$ is real. 
We consider that similar stability analysis~\cite{Hoskins:66} for frequency dependent negative resistors
(FDNR)~\cite{Zumbahlen:08,AD711,Delansky:08}
can be useful for many applications.

In the next section we will consider the work of the master equation of operational amplifiers 
applied for the purely imaginary $s=\mathrm{j}\omega.$

\section{Frequency dependent amplification for a non-inverting amplifier}
\begin{center}
\begin{figure}
\includegraphics[scale=0.3]{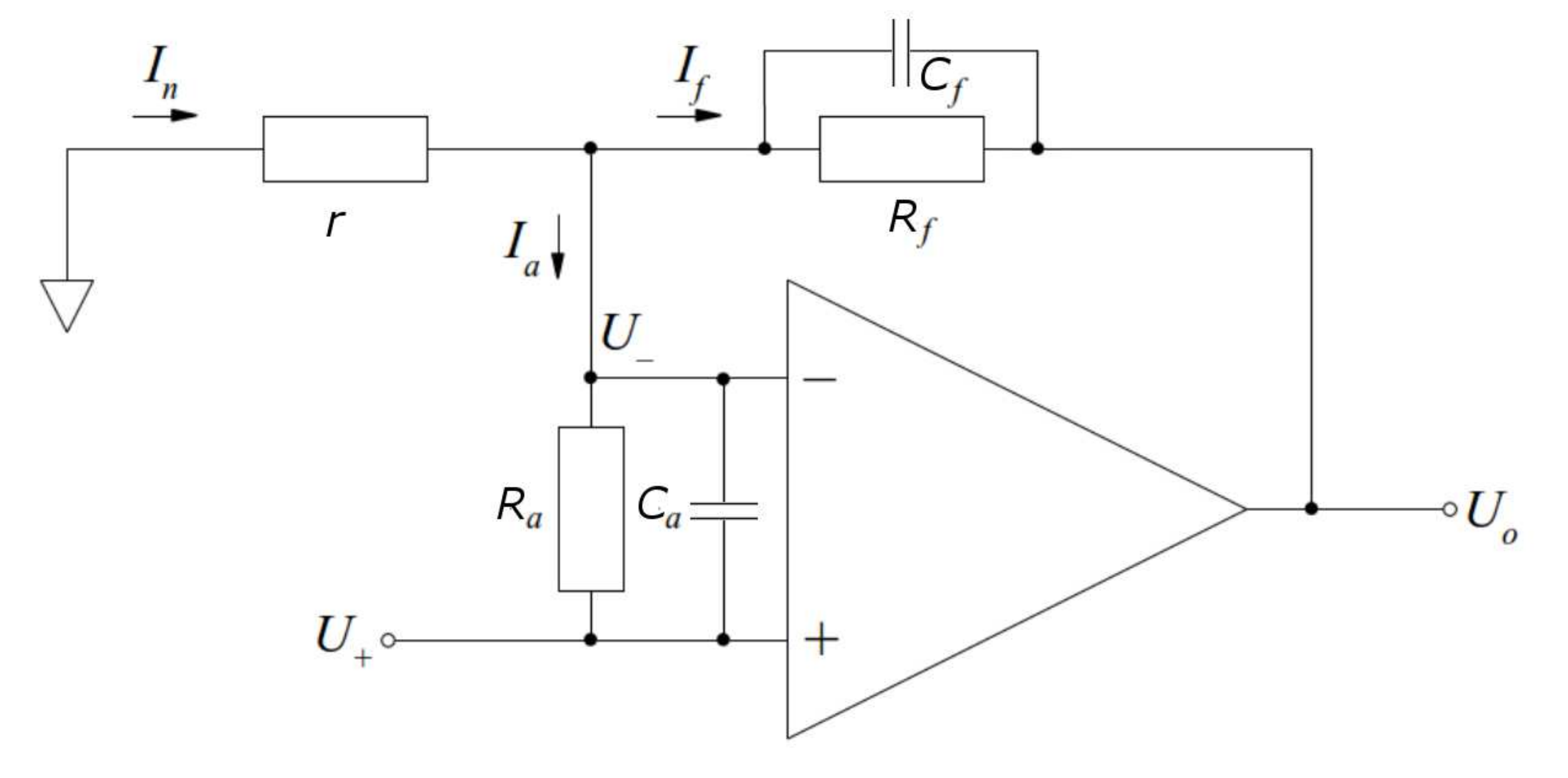}
\caption{Non-inverting amplifier with a finite input resistance
$R_a$
(non-zero input conductivity) of the operational amplifier.}
\label{fig:noninv-amp}
\end{figure}
\end{center}
For the non-inverting amplifier, depicted in Fig.~\ref{fig:noninv-amp}, the input voltage $U_i$ is applied directly to the (+) input of the operational amplifier $U_+=U_i.$
The master equation of operational amplifiers Eq.~(\ref{master}) $U_+-U_-=G^{-1}(\omega) U_0$ gives an expression for the current through the finite input impedance $z_a$
\begin{equation}
I_a = - \frac{U_+-U_-}{z_a} = -\frac{G^{-1}U_0}{z_a}.
\end{equation}
Tracing the voltage drop from $U_+$ through $z_a$ and $r \equiv z_g$ to ground gives an expression for
\begin{equation}
I_n = -\frac{U_+}{z_g} - I_a z_a = -\frac{U_+}{z_g} + \frac{G^{-1}U_0}{z_g},
\label{eq:In}
\end{equation}
where we have used the last expression for $I_a$.
Using the last two equations, the current
\begin{equation}
I_f = I_n-I_a = -\frac{U_+}{z_g} + G^{-1}U_0 \left ( \frac{1}{z_g} + \frac{1}{z_a} \right ).
\label{eq:If}
\end{equation}
Finally, tracing the voltage drop between $U_0$ through $Z_f$ and $r$ to ground, we obtain
\begin{equation}
U_0 = - I_n z_g - I_f Z_f.
\end{equation}
Substituting here Eq.~(\ref{eq:In}) for $I_n$ and Eq.~(\ref{eq:If}) for $I_f$, rearranging the terms and making convenient substitutions, we arrive at the frequency dependent amplification
\begin{eqnarray}
&&
\label{amplificationNIA}
\Upsilon_\mathrm{NIA}(\omega)\!\equiv\!\frac{U_0}{U_+}
\!=\!\frac1{Y^{-1}(\omega)+G^{-1}(\omega)[1+\epsilon_a(\omega)]},
\\&&\nonumber
Y(\omega)=\frac{Z_f(\omega)}{z_g(\omega)}+1, 
\quad z_g=r,
\quad \frac1{Z_f}=\frac1{R_f}+\mathrm{j}\omega C_f,
\\&&
\epsilon_a(\omega)=Y^{-1}r\sigma_a, \quad \sigma_a\equiv \frac{1}{z_a} \equiv
\frac1{R_a}+\mathrm{j}\omega C_a.
\end{eqnarray}

If the influence of the small feedback capacitor $C_f$
parallelly connected to the feedback resistor is negligible, i.e.for $\omega\, R_f\,C_f\ll 1$ and $G_0^{-1}\ll 1$,
we obtain the well-known formulae Eq.~(4) and Eq.~(5) of
Ref.~\cite{ADA4817},
see also Refs.~\cite{Albert:87,Jung-f:02,Franco:08,Lee:12}
\begin{eqnarray}&&
\Upsilon_\mathrm{NIA}(\omega)
=\frac{2\pi f_\mathrm{crossover}(R_f+r)}{(R_f+r)s+2\pi
f_\mathrm{crossover}\,r}, \label{eq:NIA}
\\&&\Upsilon_\mathrm{NIA}(0)=Y_0\equiv\frac{R_f}{r}+1, 
\;\, \mbox{for}\; f\!=\!\frac{\omega}{2\pi}\!\ll\!
f_\mathrm{crossover}, \nonumber
\end{eqnarray}
where the notation is once again not unique
\begin{equation}
f_\mathrm{crossover} \equiv f_0 \equiv f_{\mathrm{-3 \; dB \; open \; loop}},
\end{equation}
and here we follow the notations from the technical specifications of operational amplifiers, for example ADA4817~\cite{ADA4817}.
The Eq.~(4) of Ref.~\cite{ADA4817} coincides with Eq.~(\ref{amplificationNIA}) of the present paper
and this agreement can be considered as
implicit proof of Eq.~(\ref{eigen}) and
the time dependent master equation Eq.~(\ref{TimeDependent}).
For measurement of $f_\mathrm{crossover}$
we have to perform linear regression for 
$\left|\Upsilon_\mathrm{NIA}\right|^{-2}(\omega^2)$ using 
the experimentally determined ratio
of input to output voltages
\begin{equation}
\frac{\left| U_i\right|^{2}}{\left| U_0\right|^{2}}
=\frac1{(1+R_f/r)^2}+\frac{f^2}{f_\mathrm{crossover}^2},
\qquad f=\frac{\omega}{2\pi}.
\end{equation}
Here the first term is negligible
if the reciprocal amplification coefficient $1/(1+R_f/r)$ 
is smaller than accuracy of measurement, say 1\%.
The slope of the $\left| U_i\right|^{2}/\left| U_0\right|^{2}$ versus $f^2$ plane determines the crossover frequency and this method is invariant with respect of the used resistors.

The calculation of the pass-bandwidth requires the square of the modulus of the complex amplification, which after a straight
forward
calculation from Eq.~(\ref{amplificationNIA})
\begin{eqnarray}&&
\left|\Upsilon_\mathrm{NIA}(\omega)\right|^2\\&&\nonumber=\frac{\mathcal{N}^2(\omega)}
{\left[G_0^{-1}\mathcal{N}+ (Y_0^{-1}+\omega^2\tau_s^2)\right]^2+(\omega\tau_s)^2\left[\dfrac{\tau}{\tau_s}\mathcal{N}+I_0\right]^2},
\end{eqnarray}
where
\begin{equation}
\mathcal{N}(\omega)=1+\omega^2\tau_s^2,\quad
I_0\equiv 1-Y_0^{-1},
\quad
\tau_s\equiv \frac{C_f R_f}{Y_0}.
\end{equation}
The derivation of the frequency dependent amplification for the rest of the most widely used amplification circuits (differential \ref{Diff_Amp} and inverting amplifier \ref{Inv_Amp}) is analogous and can be found in the appendix.

\section{Determination of the crossover frequency $f_0$ of the operational amplifiers}

Let us refer to the formula for frequency dependent amplification $\Upsilon_\mathrm{NIA}(\omega)$
Eq.~(\ref{eq:NIA}),
the ratio of the output $U_0$ and input voltages $U_{\mathrm{I}}$
of a non-inverting amplifier with gain resistance $r$ and big feedback resistor $R$
taken from the specification of ADA4817~\cite{ADA4817}
and the well known monograph~\cite{Albert:87}
\begin{equation}
\Upsilon_\mathrm{NIA}(\omega)=\frac{U_0}{U_{\mathrm{I}}}=\frac{1}{Y_0^{-1}+G^{-1}(\omega)}
=\frac{1}{\left(\frac{r}{R+r} \right)+s \tau_{_0}}, 
\label{nia}
\end{equation}
where 
\begin{equation}
\quad Y_0=\frac{R}{r}+1
\end{equation}
is the static (low frequency) amplification and in order to alleviate the notations,
the index for feedback $f$ is omitted, i.e. $R_f \equiv R$.
Actually only the comparison of the two different expressions for $\Upsilon_\mathrm{NIA}(\omega)$
allows to recover the original master equation Eq.~(\ref{master})

\begin{figure}[h]
\centering
\includegraphics[scale=0.27]{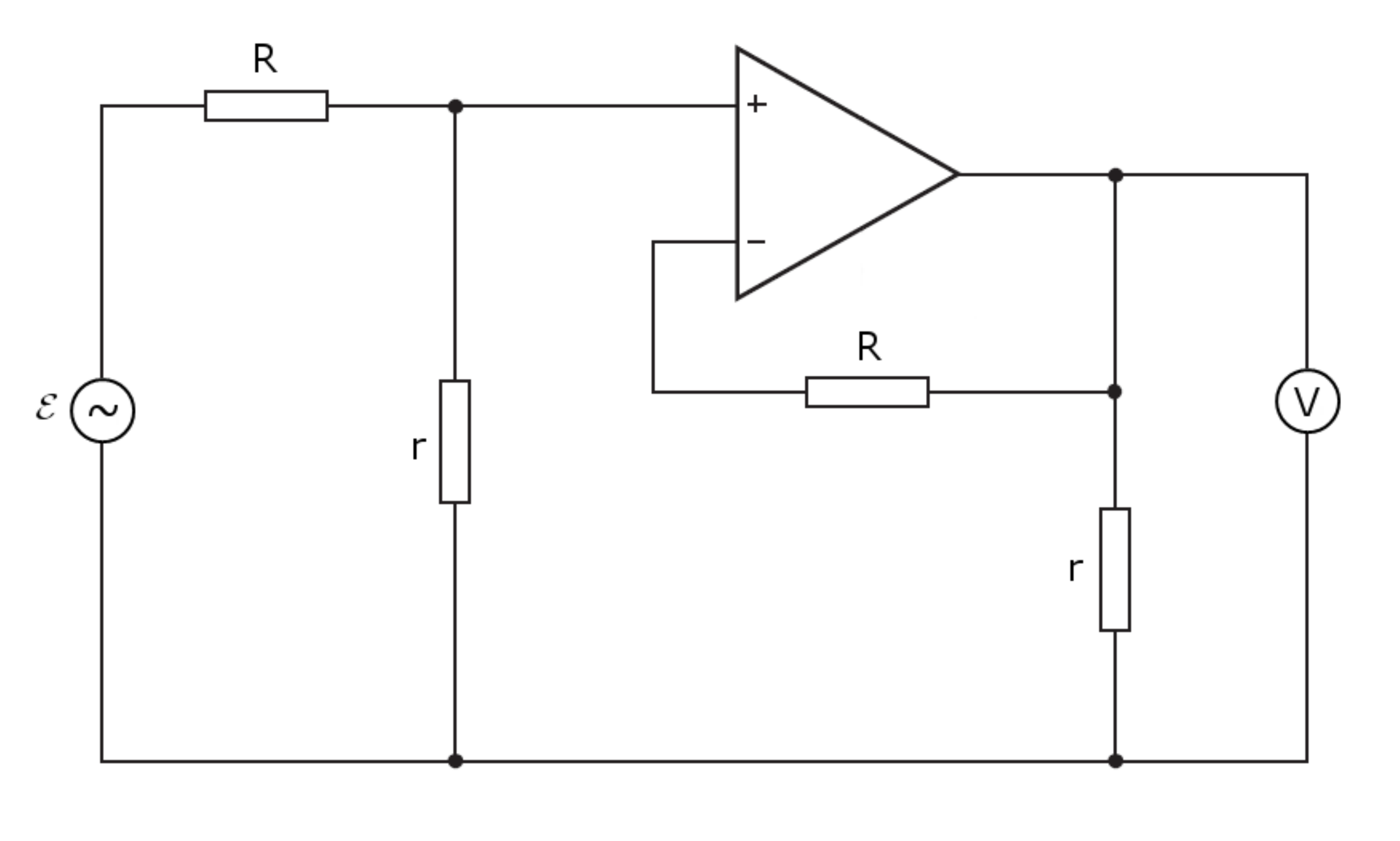}
\caption{A simple circuit for determination of the crossover frequency of an operational amplifier.
The voltage divider creates a DC (or low frequency) voltage $\mathcal{E}r/(R+r)$,
which is later restored by a non-inverting amplifier with DC amplification $Y_0=(R+r)/r.$
We change the frequency, so that at some frequency $f_\mathrm{-3dB}$ the amplification becomes two times smaller than the DC one $|Y(f_\mathrm{-3dB})|^2=Y_0^2/2$ and determine crossover frequency
as $f_0=Y_0 f_\mathrm{-3dB}.$ }
\label{f3db}
\end{figure}

When $s=\mathrm{j}\omega$ is purely imaginary for the modulus of the amplification we have
$Y(\omega)\equiv\sqrt{\left| \Upsilon_\mathrm{NIA}(\omega)\right|^2}$ and from Eq.~(1) we obtain
\begin{equation} \label{1/Y}
\frac1{Y^2(f)}=\frac{\left|U_\mathrm{I}\right|^2}{\left|U_0\right|^2}
=\frac1{Y_0^2}+\frac{f^2}{f_0^2},
\qquad f = \frac{\omega}{2 \pi}.
\end{equation}
Let us define $f_\mathrm{-3dB}$ as frequency at which the power of amplification decreases two times
\begin{equation}
Y^2(f_\mathrm{-3dB})=\frac12 Y_0^2.
\end{equation}
From Eq.~(\ref{1/Y}) we easily obtain~\cite{ADA4817}
\begin{equation}
f_0=Y_0f_\mathrm{-3dB},
\label{f0}
\end{equation}
where we have obviously considered -3 dB closed-loop gain in this expression.
A simple circuit for determination of $f_0$ using this formula is represented in Fig.~\ref{f3db}.
The voltage from the AC voltage source $\mathcal{E}$ is divided by a voltage divider
$Y_0$ times and later on at small frequencies is completely recovered by the static amplification $Y_0.$
In this way using a dozen ADA4898-2 we obtained for the crossover frequency $f_0=(46.6\pm1.3)$~MHz
for 1~V$_\mathrm{p-p}$ and $f_0=(59.9\pm 6\%)$ for 100~mV$_\mathrm{p-p}$  
which is in acceptable agreement with the values pointed out in the specification~\cite{ADA4898}.
The uncertainty is related with the used oscilloscope Hitachi V-252,
more precise measurements with Anfatec USB~Lockin~250~kHz~amplifier\cite{LockIn}
revealed that perhaps the parameter $f_0$ is temperature dependent and this requires additional research.

\section{Discussion and Conclusions}

According to the best we know, results for stability of circuits of negative impedance converters Eq.~(\ref{stab}) and Eq.~(\ref{s0}) are new results.
We analyze what the criterion for agitation of oscillations in parallel and sequential resonance circuits is.
Our research complete the well-known patent~\cite{Deboo:68}.

Using the master equation of operational amplifiers, we re-derive the frequency dependent formulae for amplification of inverting and non-inverting amplifiers given in the technical specifications~\cite{ADA4817} without any reference to contemporary books or papers.

We give a prescription for determination of the crossover frequency Eq.~(\ref{f0}) and we consider this not redundant since operational amplifiers vendors provide this parameter in their specifications, often with 50\% difference.
The exact knowledge of the crossover frequency $f_0$ is necessary when we need to precisely determine the non-ideal effects of operational amplifiers.
Even historically\cite{Ragazzini:47} in order to create electrical analogs of dynamic systems 
and to fit the differential equations which govern their dynamics, it had been necessary
to know the master equation which describes the work of the differential amplifier.

For instance, in cases when there is a need of an exact calculation of the pass bandwidth of amplifiers with active filters, the master equation Eq.~(\ref{master}) is an adequate approach.
We analyzed two typical examples of the application of this equation
(\ref{Inv_Amp} with finite internal impedance) and close the discussion 
of the formula for frequency dependent amplification coefficient
from the seminal work Eq.~(3) of Ref.~\cite{Ragazzini:47} where a circuit with the topology of inverting amplifiers (IA) has been used,
switched with opposite polarity of the operational amplifier
\begin{equation}
\Upsilon_\mathrm{IA}(s)
=-\left\{ z_\mathrm{i}(s)/Z_\mathrm{f}(s)+\left[1+z_\mathrm{i}(s)/Z_\mathrm{f}(s)\right]G(s) \right\}^{-1},
\label{inverting}
\end{equation}
where $z_\mathrm{i}$ is the gain (input $z_i \equiv z_g$) impedance and
$Z_\mathrm{f}$ is the feedback impedance switched to the operational amplifier 
with gain given by the master equation $G^{-1}\approx s\tau_{_0}$
and with infinite internal impedance.
Compare this formula with the formula for NI amplification Eq.~(7) of Ref.~\cite{ADA4817}.
As a rule classical papers are never cited in specifications and contemporary books in electronics.
But sometimes the use of classical achievements are very useful.

Let summarize the results derived in the present paper: 
1) we introduce time dependent master equation for operational amplifiers Eq.~(\ref{TimeDependent}),
2) we derive a simple method for determination of crossover frequency of operational amplifiers
as a linear regression of the experimental data processing Eq.~(\ref{1/Y}),
3) or the simpler version Eq.~(\ref{f0}).
4) We apply our method for measurement of the cross-over frequency for a contemporary low-noise 
operational amplifier ADA4898-2 and reliably observe small difference from the data given in the specification of the manufacturer~\cite{ADA4898}.
We suggest a working method for determination of the cross-over frequency
$f_0$ and systematize the corrected formulas for the main type amplifiers, see Appendix A.
5) Applying time dependent equation Eq.~(\ref{TimeDependent}), we derive stability criterion for oscillators with NIC, and obtain when NIC can excite oscillations in parallel and sequential resonance circuits.
6) As the master equation of operational amplifiers is a low frequency approximation we expect that it is good working for significant amplifications, say for gain $>5$.

Even if similar calculations are frequently found in the main textbooks about operational amplifiers, the proposed examples (NIC and non-inverting amplifier, as well as the examples in the appendix) can be useful for circuit designers and students approaching the design and the analysis of circuits based on modern operational amplifiers. 

\section*{ACKNOWLEDGMENT}
The author appreciate the cooperation with Vassil Yordanov in early stages~\cite{EPO3,arxiv} of the present research, for the friendship and stimulating comments.


\appendix


\section{Impedance derivation with operational amplifier output current}
\label{OA_oc}

\begin{figure*}[h]
\centering
\includegraphics[scale=0.33]{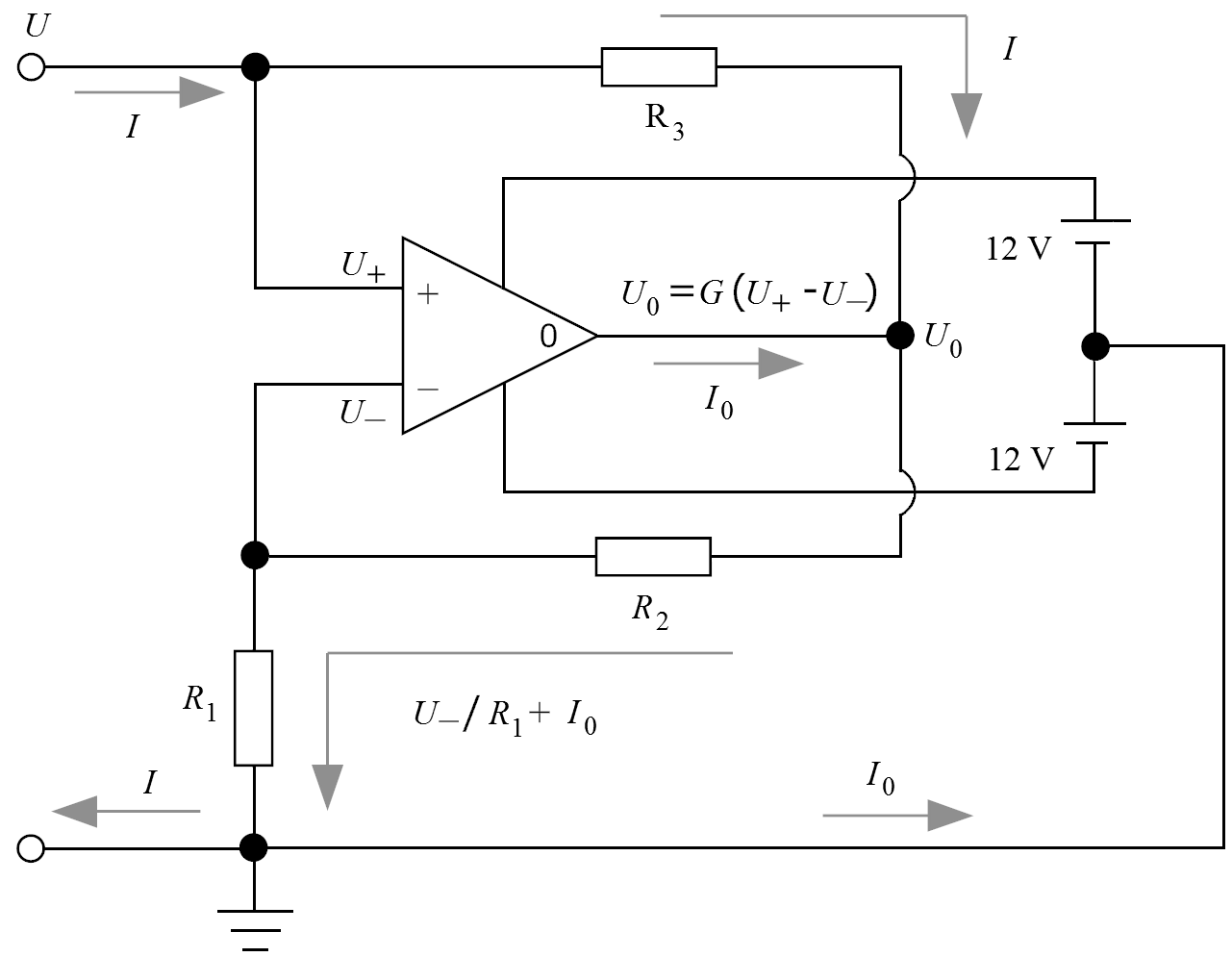}
\caption{Negative impedance converter with operational amplifier output current $I_0$ added.
Besides this, the figure is analogous to Fig.~\ref{ndc}.
As a rule currents coming from the voltage supply to the operational amplifier ($1/2 I_0$ in our case) are not shown in the circuits with operational amplifiers.
}
\label{ndc_oc}
\end{figure*}

In this appendix we are giving a detailed calculation of the problem given in the 3rd Experimental Physics Olympiad for high-school students~\cite{EPO3}.
The negative impedance converter shown in Fig.~\ref{ndc_oc} differs from the same shown in Fig.~\ref{ndc} only by the added output current $I_0$ from the operational amplifier.
Applying Ohm's law between $U_0$ and the common point, we have
\begin{equation}
U_0 = (I+I_0)(R_1+R_2),
\end{equation}
and analogously for
\begin{equation}
U_-=(I+I_0)R_1.
\end{equation}
Dividing these equations, we obtain 
\begin{equation}
U_-=U_0 \frac{R_1}{R_1+R_2 }
\end{equation}
and substitute it in Eq.(\ref{master})
\begin{equation}
U=U_0 \left ( G^{-1} + \frac{R_1}{R_1+R_2 } \right ).
\label{eq:UU}
\end{equation}
The Ohm's law applied between $U$ and $U_0$ gives
\begin{equation}
U-U_0 = I R_3,
\end{equation}
from where $U_0 = U - I R_3$ and this expression is substituted in Eq.~(\ref{eq:UU}) to obtain
\begin{equation}
U=(U - I R_3) \left ( G^{-1} + \frac{R_1}{R_1+R_2 } \right ).
\end{equation}
Therefore for the impedance of the circuit $Z=U/I$ we have
\begin{equation}
1-\frac{R_3}{Z}= \dfrac{1}{G^{-1} + \dfrac{R_1}{R_1+R_2 }},
\end{equation}
which after a little bit arrangement is
\begin{equation}
Z=\dfrac{R_3}{1-\dfrac{1}{G^{-1} + \dfrac{R_1}{R_1+R_2 }}},
\end{equation}
confirming Eq.~(\ref{NegativeImpedanceConverter}).
We have to repeat, as a rule currents coming from the voltage supplies to operational amplifiers 
and current going to the common point are not given in the circuits drawing; we are making a small exception.  

\section{Frequency dependence of the amplification for the main types of amplifiers}
\label{Y_amplifiers}

\subsection{Differential Amplifier}
\label{Diff_Amp}

\begin{figure}[h]
\begin{center}
\includegraphics[scale=0.35]{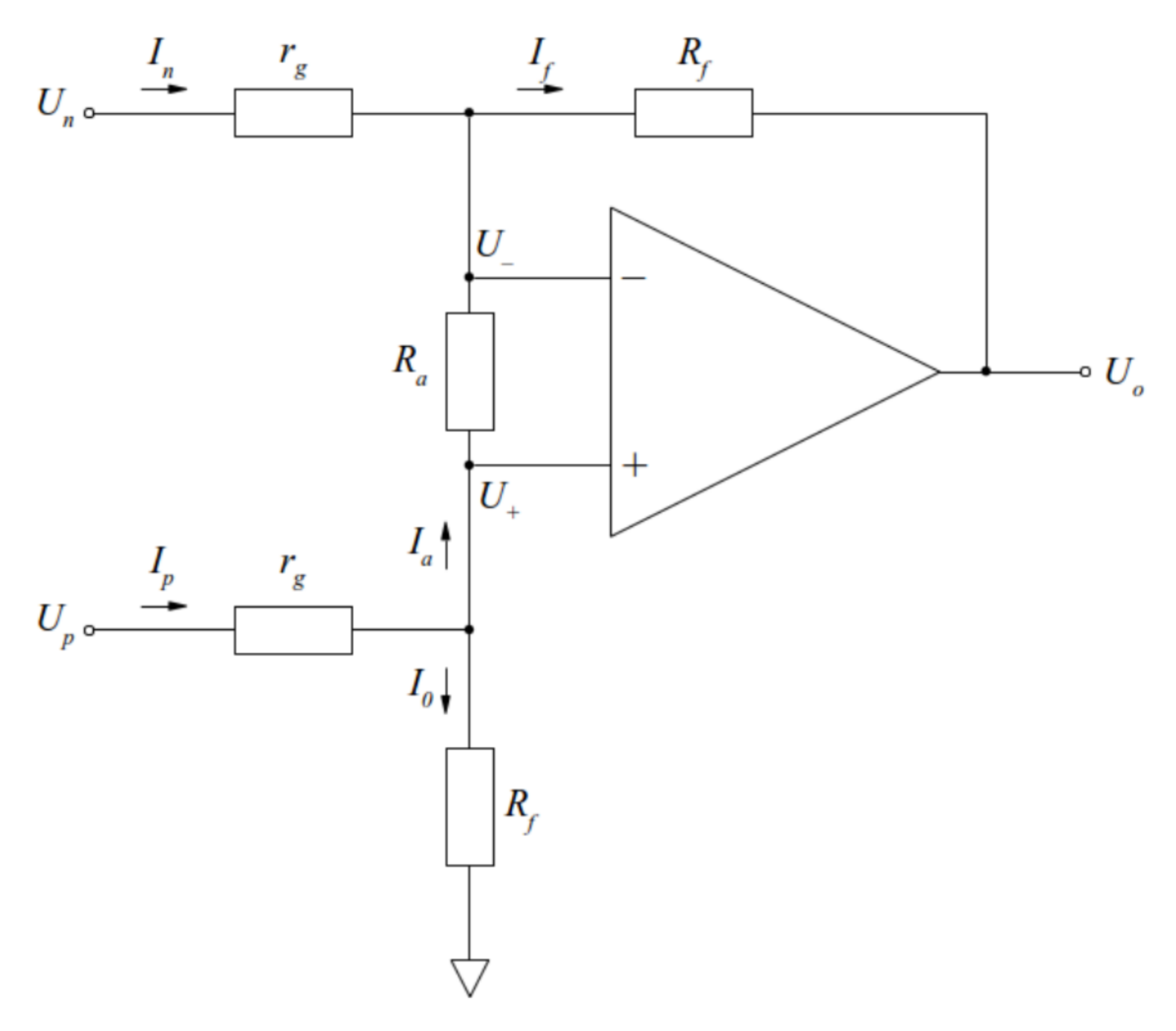}
\caption{Differential amplifier with a finite input resistance
$R_a$
(non-zero input conductivity) of the operational amplifier.} 
\label{fig:diff-amp}
\end{center}
\end{figure}
Let us trace,
see Fig.~\ref{fig:diff-amp},
the voltage drop from the negative (n) input of the differential amplifier and the output (0) voltage $U_0$.
The input current is given by the Ohm law and the voltage at the
(-) input
of the operational amplifier is given by the proportion
of the voltage divider
\begin{equation}
U_-=\left(\!I_n\!=\!\frac{U_n\!-\!U_0}{z\!+\!R_f}\right)\!R_f\!+\!U_0\!
=\!\frac{R_f}{z\!+\!R_f}U_n\!+\!\frac{z}{z\!+\!R_f}U_0.
\end{equation}
Analogous voltage divider gives the voltage of (+) input of the
operational amplifier
expressed by the voltage drop between positive (p) input of the
ground
\begin{equation}
U_+=\left(I_n=\frac{U_p-0}{z+R_f}\right)R_f+0=\frac{R_f}{z+R_f}U_n.
\end{equation}
Then we calculate the voltage difference at the inputs of the
operational amplifier
\begin{equation}
U_+-U_-=\frac{\Delta U \equiv
U_p-U_n}{z+R_f}R_f-\frac{z}{z+R_f}U_0
\end{equation}
which together with Eq.~(\ref{master})
gives
\begin{eqnarray}
\label{Delta}
&&
\Upsilon_\mathrm{\Delta}(\omega)\equiv\frac{U_0}{U_p-U_n}
\\&&\nonumber\qquad\quad
=\frac1{\Lambda(\omega)+\left[\Lambda(\omega)+2\epsilon_a(\omega)\right]
G^{-1}(\omega)+G^{-1}(\omega)},
\\&&\nonumber
\Lambda(\omega)=\frac{z_g(\omega)}{Z_f(\omega)}, 
\quad z_g=r+\frac1{\mathrm{j}\omega C_g},
\quad Z_f=R_f
\\&&\nonumber
\varepsilon_a(\omega)=z_g(\omega)\sigma_a(\omega), \quad
\sigma_a(\omega)\equiv \frac1{R_a}+\mathrm{j}\omega C_a,
\end{eqnarray}
cf. Eq.~(3.14) of Ref.~\cite{Albert:87}.
Here we added the correction $\varepsilon_a$ of internal
conductivity $\sigma_a$ between inputs of operational amplifier in differential mode.
A detailed derivation will be given later in Appendix~B.

An instrumentation amplifier can be considered as a buffer of two symmetric non-inverting amplifiers
followed by a difference amplifier.
The amplification of the instrumentation amplifier is just the product of the amplifications of
the non-inverting and the difference amplifiers
\begin{equation}
\Upsilon_\mathrm{INS}(\omega)=\Upsilon_\mathrm{NIA}(\omega) \Upsilon_\Delta(\omega).
\end{equation}

\subsection{Inverting Amplifier}
\label{Inv_Amp}

\begin{center}
\begin{figure}[h]
\includegraphics[scale=0.25]{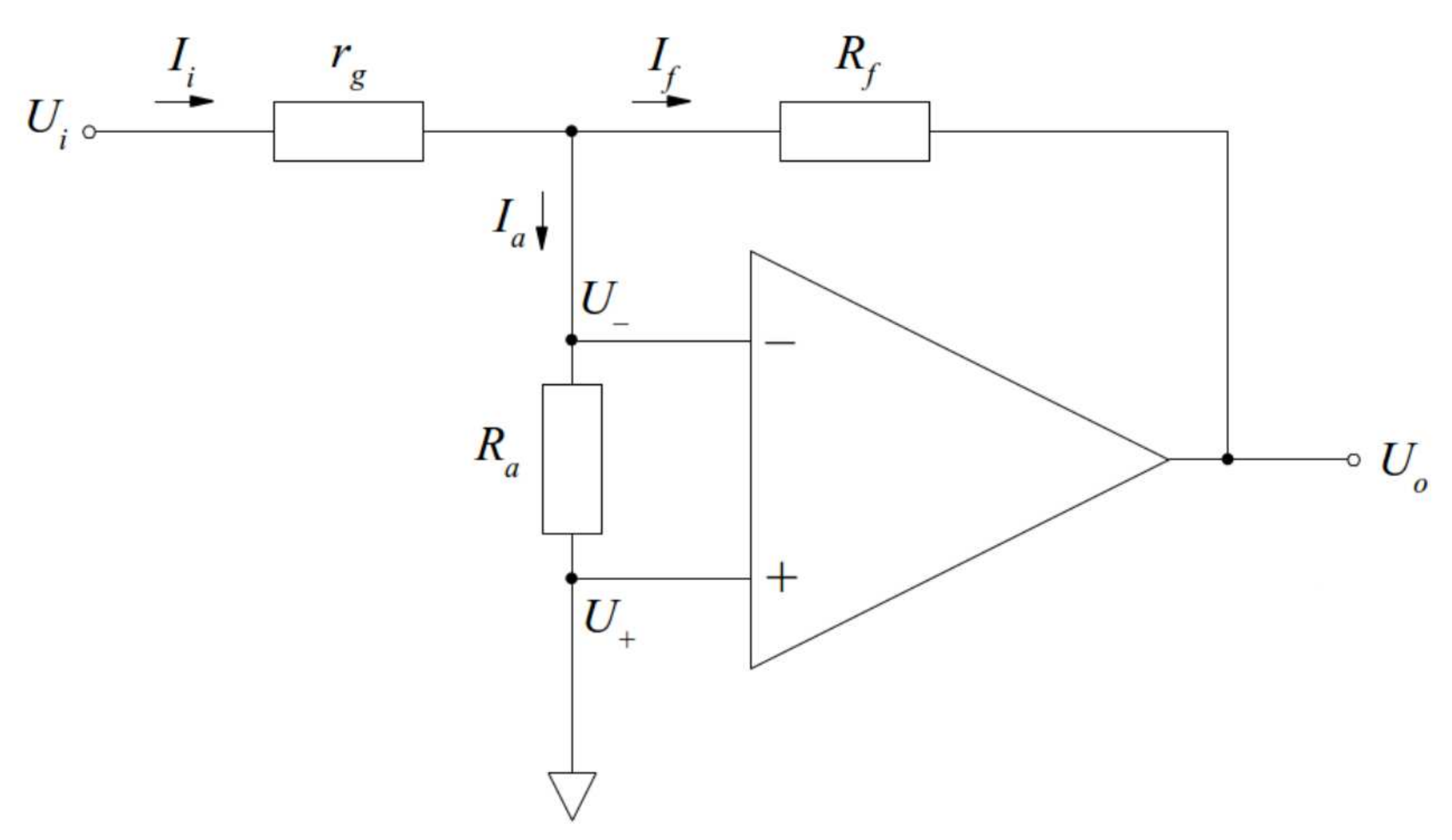}
\caption{Inverting amplifier with a finite common mode input
resistance $R_a$
(non-zero input conductivity) of the operational amplifier.} 
\label{fig:inv-amp}
\end{figure}
\end{center}
If for the differential amplifier,  
see Fig.~\ref{fig:inv-amp},
the positive (p) input is grounded 
$U_p=0$,  this leads to $U_+=0$ and $\Delta U=-U_n$ 
hence for the transmission function we get 
$\Upsilon_\mathrm{IA}(\omega)=-\Upsilon_\mathrm{\Delta}(\omega)$or finally 
\begin{eqnarray}
&&
\Upsilon_\mathrm{IA}(\omega)\equiv\frac{U_0}{U_n}
\\&&\nonumber\qquad\quad
=-\frac1{\Lambda(\omega)+\left[\Lambda(\omega)+\epsilon_a(\omega)\right]
G^{-1}(\omega)+G^{-1}(\omega)},
\\&&\nonumber
\Lambda(\omega)=\frac{z_g(\omega)}{Z_f(\omega)}, 
\quad z_g=r+\frac1{\mathrm{j}\omega C_g},
\quad Z_f=R_f
\\&&\nonumber
\varepsilon_a(\omega)=z_g(\omega)\sigma_a(\omega), \quad
\sigma_a(\omega)\equiv \frac1{R_a}+\mathrm{j}\omega C_a.
\end{eqnarray}
The subtle difference with Eq.~(\ref{Delta}) is only the
coefficient in front of $\epsilon_a$,
cf. Eq.~(3.10) of Ref.~\cite{Albert:87}.
For negligible the internal conductivity $\sigma_a$ we derive
\begin{eqnarray}
&&
\label{IA}
\Upsilon_\mathrm{IA}(\omega)
=-\frac1{\Lambda(\omega)+\left[1+\Lambda(\omega)\right]G^{-1}(\omega)},
\end{eqnarray}
i.e. Eq.~(\ref{inverting}) which 
A large capacitance of the gain impedance stops the offset
voltages coming
from former amplification blocks.  
Additionally, for high enough working frequencies 
$\omega\, r\,C_g\gg 1$ and 
negligible static-inverse-open-loop gain
$G_0^{-1}\ll 1$
we obtain the well-known formulae Eq.~(7) and Eq.~(8) of
Ref.~\cite{ADA4817},
see also Ref.~\cite{Albert:87}.
Actually coincidence of Eq.~(7) ofRef.~\cite{ADA4817} with 
Eq.~(\ref{inverting}) from our paper with reverse engineering gives a proof 
of our time dependent master equation Eq.~(\ref{TimeDependent}).
For frequencies much lower than the crossover frequency 
$f_\mathrm{crossover}$ of the operational amplifier
the amplification is in a wide band frequency independent
\begin{eqnarray}&&
\Upsilon_\mathrm{IA}(\omega)\approx-\frac{2\pi
f_\mathrm{crossover}R_f}
{(R_f+r)s+2\pi f_\mathrm{crossover}\,r},
\\&&\Upsilon_\mathrm{IA}=-\frac{R_f}{r}
\qquad \mbox{for}\quad f=\frac{\omega}{2\pi}\ll
f_\mathrm{crossover}.
\end{eqnarray}
For calculation of the pass-bandwidth we will need to know the 
square of the modulus of the complex amplification 
\begin{eqnarray}&&
\left|\Upsilon_\mathrm{IA}(\omega)\right|^2=\left|\Upsilon_\Delta(\omega)\right|^2
\\&&\nonumber=\frac{(\omega\tau_g)^2}
{\left[1\!+\!G_0^{-1}\!-\!M\omega^2\tau\tau_g\right]^2
+(\omega\tau_g)^2\left[\dfrac{\tau}{\tau_g}\!+\!\Lambda_0\!+\!G_0^{-1}M\right]^2},
\\&&\nonumber\quad
\Lambda_0\equiv\frac{r}{R_f},\qquad M\equiv 1+\Lambda_0,\qquad
\tau_g\equiv C_g R_f.
\end{eqnarray}

For large enough amplification $r\ll R_f$ from Eq.~(\ref{IA}) we
obtain
\begin{eqnarray}&&\nonumber
\frac{R_f}{r}\gg 1,\quad \omega r C_g\gg 1, \quad \omega
\tau\ll1,\\&&
\Upsilon_\mathrm{IA}(\omega)
\approx
-\frac{R_f}{r+\frac{1}{\mathrm{j}\omega
C_g}+\mathrm{j}L_\mathrm{eff}\omega},
\\&&\nonumber
L_\mathrm{eff}\equiv \tau R_f,
\quad Q=\frac{\sqrt{L_\mathrm{eff}/C_g}}{r} =\frac{\sqrt{\tau
R_f/C_g}}{r} \ll1,
\end{eqnarray}
and in this approximation the frequency dependence of the
amplification
reminds of the frequency response of an overdamped oscillator with
low quality factor $Q$.

\section{Influence of finite common mode input conductivity
of operational amplifiers}
\label{InternalConductivity}

In this appendix we will investigate the influence of 
a small conductivity between the inputs of an operational
amplifier in differential mode.
For the simplicity of the notations we will write 
only the big differential mode input resistance $R_a$, 
while above we have included the
differential mode input capacitance.
In the next subsection we will start with the inverting
amplifier.

\subsection{Inverting amplifier}

Standard scheme of the inverting amplifier is presented in Fig.~\ref{fig:inv-amp}.
Let us write the Kirchhoff equations. 
The input current $I_i$ is branching 
to the current $I_f$ passing trough the feedback resistor $R_f$
and the current $I_a$ flowing between (-) and (+) inputs of the
operational amplifier
\begin{equation}
\label{inv:branching}
I_i =I_a+ I_f .
\end{equation}
Let's trace the voltage drop between the input voltage by gain
resistor $r_g$ and input resistor $R_a$
to the ground
\begin{equation}
\label{inv:ground}
U_i = R_a I_a  +  r_g I_i.
\end{equation}
The Ohm law of the internal resistor of the operational amplifier is
\begin{equation}
\label{inv:U+-}
-R_a I_a= U_+-U_-=G^{-1}U_o,
\end{equation}
where the potential difference between inputs voltages 
is expressed by the output voltage $U_o$ and reciprocal open
loop gain $G^{-1}.$
Finally starting from the output voltage $U_o$ and passing
trough the feedback resistor $R_f$
and gain resistor $r_g$ we reach the input voltage $U_i.$
\begin{equation}
\label{inv:output}
U_i =U_o +  R_f I_f + r_g I_i.
\end{equation}
From Eq.~(\ref{inv:U+-}) we express $I_a=-U_o/R_aG$
and substitute in Eq.~(\ref{inv:ground}) which now reads
\begin{equation}
U_i=-G^{-1}U_o+r_gI_i,
\end{equation}
and gives
\begin{equation}
I_i=\left(U_i+G^{-1}U_o\right)/r_g.
\end{equation}
The substitution of $I_i$ in Eq.~(\ref{inv:branching}) 
gives 
\begin{equation}
I_f=\frac{U_i}{r_g}+\left(\frac1{r_g}+\frac1{R_a}\right)\frac{U_o}{G}.
\end{equation}
Substitution of derived in such a way currents in
Eq.~(\ref{inv:output}) gives for the output voltage $U_o$ the
well-known result Eq.~(3.10)
of Ref.~\cite{Albert:87}
\begin{equation}
\frac{U_o}{U_i} = - \frac{R_f/r_g}{1+G^{-1} \left ( 1+
\dfrac{R_f}{r_g} + \dfrac{R_f}{R_a} \right )}.
\label{inv:fin}
\end{equation}

In the next subsection we will perform analogous consideration
for the non-inverting amplifier.

\subsection{Non-inverting amplifier}

The non-inverting amplifier circuit is depicted in Fig.~\ref{fig:noninv-amp}.
The current from the ground $I_n$ is branching to
the current $I_f$ passing through the feedback resistor $R_f$
and
the current $I_a$ flowing through the operational amplifier (-) and (+) inputs.
The charge conservation in the branching point is
\begin{equation}
\label{noninv:branching}
I_n = I_f + I_a.
\end{equation}
Tracing the voltage drop from ground through the gain resistor
$r_g$ and
the operational amplifier internal resistance $R_a$ to the input voltage $U_i$
we have
\begin{equation}
\label{noninv:ground}
U_+=-r_g I_n-R_a I_a.
\end{equation}
Where we take into account that the input voltage is directly
applied to
the positive input of the operational amplifier
where $U_i \equiv U_+$.
The Ohm law of the internal resistor of the operational amplifier is
\begin{equation}
\label{noninv:U+-}
-R_a I_a = U_+-U_-=G^{-1} U_o,
\end{equation}
where the potential difference between the input voltage 
is expressed by the output voltage $U_o$ and reciprocal open
loop gain $G^{-1}.$
The last circuit to consider is the voltage drop from the output
voltage $U_o$
through $R_f$ and $r_g$ to ground
\begin{equation}
\label{noninv:Uo}
U_o = -I_f R_f - I_n r_g.
\end{equation}
Expressing the currents $I_f$ from Eq.~(\ref{noninv:branching}),
$I_n$ from Eq.~(\ref{noninv:ground}),
$I_a$ from Eq.~(\ref{noninv:U+-}) and substituting them in
Eq.~(\ref{noninv:Uo}),
for the ratio of the output voltage $U_o$ to the input voltage
$U_i$ we obtain
\begin{equation}
\frac{U_o}{U_i} = \frac{R_f/r_g+1}{1+G^{-1} \left ( 1+
\dfrac{R_f}{r_g} + \dfrac{R_f}{R_a} \right )}.
\end{equation}
The differential amplifier analyzed in the next subsection is
slightly more complicated.

\subsection{Differential amplifier}

The differential amplifier circuit is shown in Fig.~\ref{fig:diff-amp}.
The feedback current $I_f$ going through the feedback resistor
$R_f$ is a sum of
the input current $I_n$ from the input voltage $U_n$ and
the current $I_a$ flowing through the operational amplifier (+) and (-) inputs
\begin{equation}
\label{diff:nbranching}
I_f = I_n + I_a.
\end{equation}
The other input current $I_p$ from the input voltage $U_p$
flowing through the
gain resistor $r_g$ is branching to the current $I_a$ and
the current $I_0$ flowing through the other feedback resistor
$R_f$ to ground
\begin{equation}
\label{diff:pbranching}
I_p= I_0 + I_a.
\end{equation}
Now let trace the circuit between $U_n$ and the output voltage
$U_o$
passing through $r_g$ and the feedback resistor $R_f$
\begin{equation}
\label{diff:Un}
U_n - U_o = r_g I_n  + R_f  I_f .
\end{equation}
Another circuit to consider is from $U_p$ through $r_g$ and
$R_f$ to ground,
for which the Kirchhoff's voltage law reads
\begin{equation}
\label{diff:Up}
U_p = r_g I_p + R_f  I_0.
\end{equation}
Substituting $I_f$ from Eq.~(\ref{diff:nbranching}) into
Eq.~(\ref{diff:Un}), for $I_n$ we derive
\begin{equation}
\label{diff:In}
 I_n = \frac{U_n-U_o}{r_g+R_f}-\frac{R_f}{r_g+R_f} I_a.
\end{equation}
Expressing $I_0$ from Eq.~(\ref{diff:pbranching}) and
substituting it into Eq.~(\ref{diff:Up}),
for $I_p$ we obtain
\begin{equation}
\label{diff:Ip}
 I_p = \frac{U_p}{r_g+R_f}+\frac{R_f}{r_g+R_f} I_a.
\end{equation}
The difference between positive  $U_+$ and negative $U_-$
input of the operational amplifier can be expressed
simultaneously
by the output voltage $U_o$ and the Ohm law for the internal
resistance of
the operational amplifier
\begin{equation}
\label{diff:U+-}
R_a I_a =  U_+ - U_- = G^{-1} U_o.
\end{equation}
The current $I_a=U_o/GR_a$ is easily expressed by the open loop
gain $G$,
output voltage $U_o$ and internal resistance of the operational amplifier in
differential mode $R_a$.
Finally we write down the Kirchhoff's voltage law for the
circuit
from $U_p$ to $U_n$ through the resistors with values 
$r_g$, $R_a$ and $r_g$
\begin{equation}
\label{diff:dU}
U_p-U_n \equiv \Delta U =  r_g I_p + R_a I_a  - r_g I_n,
\end{equation}
where $\Delta U = U_p-U_n$ denotes the input voltage difference
between positive $U_p$
and negative $U_n$ input voltages.
Substitution of the already derived expressions for the input
currents
$I_p$, $I_n$ and the internal operational amplifier current $I_a$ into
Eq.~(\ref{diff:dU})
yields for the transmission coefficient, i.e.
ratio of the output voltage $U_o$ 
and input voltage difference $\Delta U$,
\begin{equation}
\frac{U_o}{\Delta U} = \frac{R_f/r_g}{1+G^{-1} \left ( 1+
\dfrac{R_f}{r_g} + 2\dfrac{R_f}{R_a} \right )}.
\end{equation}

For high frequencies we have to take into account also the input
capacitance of the operational amplifier in differential mode
$R_a\rightarrow Z_a=R_a/(1+\mathrm{j}\omega R_a C_a).$
For the used ADA4898\cite{ADA4898} $R_a=5\,\mathrm{k}\Omega$ and $C_a=2.5\,\mathrm{pF}.$
In all formulae we consider frequency dependent open-loop gain 
$G^{-1}(\omega)=G^{-1}_0+\mathrm{j}\omega\tau$
given by the master equation of the operational amplifiers
Eq.~(\ref{master}).


\end{document}